# DIRECT OBSERVATION OF VORTEX SHELLS AND MAGIC NUMBERS IN MESOSCOPIC SUPERCONDUCTING DISKS

I.V. Grigorieva[1*], L.Y. Vinnikov[1,2], W. Escoffier[1], J. Richardson[1], S.V. Dubonos[3], V. Oboznov[2], A.K. Geim[1]

[1]School of Physics and Astronomy, University of Manchester, Oxford Rd., Manchester M13 9PL, United Kingdom

[2]Institute of Solid State Physics, Russian Academy of Sciences, Chernogolovka, 142432, Russia

[3]Institute of Microelectronics Technology, Russian Academy of Sciences, Chernogolovka, 142432, Russia

## ABSTRACT

We have visualised and studied vortex configurations in mesoscopic superconducting disks using the Bitter decoration technique. For a broad range of vorticities $L$ the circular geometry is found to result in the formation of concentric shells of vortices. From images obtained on disks of different sizes in a range of magnetic fields we were able to trace the evolution of vortex states and identify stable and metastable configurations. Furthermore, quantitative analysis of shell filling with increasing $L$ allowed us to identify 'magic' numbers corresponding to the appearance of consecutive new shells.

---

[*] To whom correspondence should be addressed

The properties of a superconductor are known to change dramatically as its size becomes comparable with the size of Cooper pairs, i.e., the superconducting coherence length $\xi$ (for a recent review, see ref. [1]). When placed in a magnetic field, such a superconductor (usually referred to as mesoscopic) will reside in one of a series of discrete states satisfying the quantization conditions, with each state characterised by a vorticity $L$ (the number of fluxoids in the sample). The behaviour of mesoscopic superconductors has attracted much interest, especially since the prediction of two fundamentally different states of vortices in them – the multi-vortex state with singly quantised vortices [2-6], expected over most of the magnetic field range below the second critical field $H_{c2}$, and a giant-vortex state with the same vorticity $L$ [7,8]. This has prompted a large number of numerical and experimental studies that continue to bring surprises: the symmetry-induced antivortices [9], paramagnetic Meissner effect [10], first evidence for giant vortex states [11], multiple phase transitions within the superconducting state [11,12], negative and fractional vortices [13] are just some of the examples.

On the other hand, vortices in small superconductors are a typical representative of interacting particles in finite systems (other examples are, e.g., electrons in artificial 'atoms' [14] and vortices in rotating superfluids [15-17]). The details of particle configurations in such systems, e.g., stability of different states or the rules of formation of consecutive shells of particles, are governed by the rich and subtle physics arising due to the competition between the effects of particle interactions and confinement [6,14,15]. As vortices in superconductors are relatively easy to visualise, understanding and especially direct imaging of the vortex state should help understanding of confined interacting systems in general. Recent numerical studies of vortices in confined geometries considered mostly thin circular disks, either very small (radius $R$=4 to 6$\xi$) with a maximum number of singly quantised vortices $L$=11 [3-6] or relatively large (radius $R$=50$\xi$, containing over 200 vortices) where formation of the triangular lattice becomes a dominant feature [18]. However, none of the predicted vortex configurations has been observed directly so far, and the properties of the multi-vortex state for intermediate vorticities 10<$L$<100 remain largely unknown. The last few years have seen several groups attempting to visualise vortices in confined geometries but no results have been reported so far. The only exception is a recent observation using a scanning SQUID microscope [19], which was limited to a maximum of 3 vortices in relatively large square and triangular samples (30-50$\mu$m).

In this report, we present the first direct observations of stationary vortex states in small superconducting disks for $L$ =0 to 40. Well-defined shell structures were found over the entire vorticity range, and quantitative analysis of the evolution of vortex states with increasing $L$ allowed us to identify the rules of shell filling and 'magic' numbers corresponding to the appearance of consecutive new shells. Only one vortex configuration was found for each $L$ at small vorticities $L \leq 4$ but up to four different states for each $L$ were observed for larger vorticities. Using statistical analysis of the images obtained on hundreds of samples we were able to distinguish between stable and metastable states for certain $L$.

The mesoscopic samples for this study were made from a 150 nm thick Nb film deposited on a Si substrate using magnetron sputtering. The film's superconducting parameters were: transition temperature $T_c$ =9.1K, magnetic field penetration depth $\lambda(0) \approx$ 90 nm; coherence length $\xi(0) \approx$ 15nm; upper critical field $H_{c2}(0) \approx$ 1.5 T. The details of the Bitter decoration technique used to visualise individual vortices are described elsewhere [20]. Fig. 1a shows a typical vortex structure in the starting macroscopic film. The vortex arrangement is uniform but disordered, which is a signature of the presence of a large number of small pinning centres, typical for sputtered Nb films [21]. Using e-beam lithography and dry etching, these films were made into arrays of small "dots" – circular disks, triangles and squares of 4 different sizes - $\approx$1μm, $\approx$ 2μm, $\approx$3μm and $\approx$5μm (see Fig. 1b). A whole array, containing over 500 dots, was decorated in each experiment, allowing us to obtain a snapshot of up to a hundred vortex configurations in dots of the same shape and size, produced in identical conditions (magnetic field $H_{ext}$, temperature $T$, pressure, etc.). It was therefore possible not only to simultaneously visualise vortex configurations in samples of different sizes but also to gain enough statistics for quantitative analysis of the observed vortex states in terms of their stability, sensitivity to shape irregularities, and so on. Below we present the results obtained after field-cooling to $T \approx$ 1.8 K in external fields ranging from 20 to 160 Oe. Using many different combinations of the external field and dot sizes allowed us to achieve vortex configurations with almost all possible vorticities $L$ up to 50. In the rest of this paper we will concentrate on the results obtained for circular disks only. Vortex states in more complex geometries (squares and triangles) will be discussed elsewhere.

Fig. 1*c* shows vortex configurations observed in the same experiment on disks of three different diameters. It demonstrates the excellent resolution of our technique, as well as the main features of the observed vortex configurations. A typical size of a vortex 'image' in Fig. 1*c* is ~0.25-0.3μm while a typical distance between neighbouring vortices is *a*~0.8-1 μm, thus ensuring that vortex positions could be determined easily and unambiguously. For *L*=1 to 4, the vortices form rather obvious symmetric configurations (the states with *L*=2 and *L*=3 are shown in the figure). As the vorticity increases above 4, vortices start forming well defined shells, e.g., two shells in 3μm disks of Fig. 1*c* and three shells in 5 μm disks. Using the standard notation for configurations of particles in confined geometries [4-6,14,15], we denote the states of Fig. 1*c* as (2), (3), (2,7), (2,8), (2,7,13) and (3,8,13). The alignment of vortices in different shells seen in these images is typical of vortex configurations observed in all our experiments. In most cases, vortices in the inner shell tended to align with the gaps between vortices in the outer shell. This is evident, for example, for the states (2,7) and (2,8) in Fig. 1*c*. As the difference between the vortex numbers in neighbouring shells becomes smaller, this tendency becomes less apparent, presumably overridden by the requirement of a uniform distribution of vortices within a shell [see, e.g. state (2,7,13) in Fig. 1*c*].

With reference to the vortex structure in the unpatterned macroscopic film, it is clear that the effect of confinement dominates over pinning. In circular geometry this was found to be the case for all vorticities up to $L \approx 40$. For much larger $L > 80$, such that a triangular lattice is expected to form in the disk's centre, with just one or two circular shells near the edge [18], we observed disordered vortex arrangements reminiscent of that in the parent film, i.e., pinning becomes dominant again for relatively large disks.

Using the field-cooling regime allowed us to achieve vortex configurations close to the stable (lowest-energy) states, but not excluding the metastable states. Indeed, in a confined geometry, magnetization $M=<B>-H$ as a function of the external magnetic field *H* is represented by a family of curves *M*(*H*) corresponding to different vortex states (i.e., different vorticities *L*, as well as different vortex configurations for the same *L* [10,12,22]). Each state can be realised over a range of magnetic fields (up to 20-30G for a superconductor of a few μm size) and several different states can exist for any given value of the applied field, but only the state with the most negative magnetisation is thermodynamically stable. In the absence of external disturbances, field-cooling results in a vortex state with a smaller magnetisation than that corresponding to the thermodynamically stable state for a given $L$. This happens because thermodynamically stable states near $T_c$, which correspond to the surface critical field $H_{c3}$, require a flux greater than $L\Phi_0$ ($\Phi_0$ is the flux quantum) and this flux can be trapped in low-temperature states due to the presence of the edge [10]. However, as demonstrated in ref. [10], this 'supercooled' state can be relaxed to a state closer to the thermodynamic equilibrium by an external disturbance. In decoration experiments, this role is played by additional heat inevitably introduced into the system during thermal evaporation of Fe particles [20]. The relaxation towards equilibrium is further aided by roughness of the sample boundary [10, 22].

To demonstrate the relationship between the vorticity and the magnetic flux through the disk area, $\Phi = H_{ext} \cdot S$ (where $S$ is the sample area), Figure 2 shows the results for the entire vorticity range, obtained in different experiments. Here data points correspond to measurements on individual disks, rather than the average values for different experiments and disk sizes. For all vorticities, the flux required to form $L$ vortices in a disk is considerably larger than for a same-area part of a macroscopic film, $\Phi_{film} = \Phi_0 L$ – c.f. straight solid line in Fig. 2. This result becomes clearer when plotted as an excess flux $\delta\Phi$ (compared to the 'macroscopic film') needed to add one vortex to a state $L$-1 [23] - see inset in Fig. 2. Indeed, a total flux of about $3.5\Phi_0$ is required for the formation of a first vortex, about $2.8\Phi_0$ for the addition of a second one and so on. Accordingly, the states with small $L$ ($L<5$) are stable over appreciable intervals of magnetic field $\Delta H$ even for relatively large disks: for example, the state $L=1$ in a 2μm disk is stable over $\Delta H \approx 20$ Oe, the state $L=2$ over $\Delta H \approx 10$ Oe, etc. At $L>10$ the excess flux saturates at $\delta\Phi \approx 0.2\Phi_0$, which can be explained by finite demagnetisation even in the limit of large $L$. The observed behavior is in good agreement with results of the numerical analysis based on nonlinear Ginzburg-Landau equations, as shown by a fit (dashed line) to the $\delta\Phi(L)$ dependence found in ref. [5].

To check whether the flux expulsion results of Fig. 2 depend on the strength of the external field, we have chosen several different values of $H_{ext}$ such that they produced the same $L$ states. For example, the state $L=2$ was realised in disks with diameter $d \approx 2.2$μm at $H_{ext}=40$ Oe and in disks with $d \approx 1$μm at $H_{ext}=160$ Oe; the state $L=21$ in disks with $d \approx 4.9$μm at $H_{ext}=40$ Oe and in disks with $d≈2.4\mu m$ at $H_{ext}=160$ Oe, etc. No difference was found with our experimental accuracy within the above field range.

For small vorticities ($L≤4$), only one vortex configuration for a given $L$ was found in all experiments – one vortex sitting in the centre for $L=1$ and similarly obvious symmetric arrangements for $L=2-4$ (see Fig. 1 and images in Fig. 3). For larger $L$, however, at least two or more vortex configurations were found in any given experiment (i.e. in the same external field and under identical decoration conditions). This is demonstrated by the histogram in Fig. 3a, where, firstly, two different vorticities ($L=9$ and $L=10$) and, secondly, three distinct states for each vorticity have been observed in the same experiment for disks of approximately the same size. The two different vorticities can be explained by variations in size and shape of individual disks, as these were found to lead to significant variations in the actual total flux, $\Phi$ (up to $\Phi_0$ in larger samples). However, these variations alone cannot account for the observed multiplicity of vortex states for the same $L$. Indeed, different states were found in the same experiment even for disks with the same area, smooth boundaries and minimal deviations from the ideal circular shape (see for example the images of states (2,8) and (3,7) in Fig. 3a).

The only possible explanation is that we observe not only the stable (lowest-energy) state but also metastable states, corresponding to local energy minima, which differ among themselves only slightly in free energy [4-6, 8]. Indeed, in all experiments one of the configurations for a given $L$ was observed typically twice more frequently than the others (see Fig. 3*a*), indicating that it has a lower energy and/or a larger interval of stability [5,6]. We therefore identify this configuration as the stable (ground) state, with the others representing metastable states for a given $L$. The results of such analysis are summarised in Table I. Most of the stable states found experimentally agree with the results of numerical simulations [3-6] (states for $L=0$ to 5, $L=7$ and $L=8$). However, the predicted ground state (6) for $L=6$ (all 6 vortices in one shell, as shown in Fig. 3*b*) was found in just 5% of cases, indicating that it is much less stable than the state (1,5) with one vortex in the centre. Similarly, for $L=9$ the predicted ground state (1,8) was found in just a few cases, mostly with a somewhat square – rather than circular – symmetry (see Fig. 3*a*). The states (1,9) and (2,9) were never seen in our experiments. This discrepancy can be due to the fact that our samples are much larger than those considered numerically: a typical disk radius in our experiments $R≈100\xi$ as compared with e.g. $R=6\xi$ in ref. [6]. Indeed, the vortex configurations observed in our experiments are in better agreement with those predicted [15] and found [16] for rotating liquid helium or for a finite system of charged particles [14], where the system size was much larger than the characteristic size of a particle, i.e., similar to our case. From a different perspective, the agreement between our observations for vortices in superconductors and predictions for other confined systems provide yet another proof of similarity in their behaviour.

The square-symmetry states $(1,8)_\diamond$ and $(2,8)_\diamond$ shown in Fig. 3*a* are notable exceptions from the general rule where, as expected for the circular symmetry of the boundary, vortex positions fall very closely into concentric rings. In the state $(1,8)_\diamond$ the 8 outer vortices sit on the perimeter of a square, rather than on a ring, effectively having the same four-fold symmetry as the state *L*=4 (see Fig. 3*b*). We do not believe that this effect is due to pinning of one of the vortices. Indeed, the state $(1,8)_\diamond$ was observed more frequently than the circular state (1,8), while the state (2,8) was found predominantly with the circular symmetry, presumably because the two vortices in the centre break the four-fold symmetry. One might expect to find the square symmetry also for larger vorticities matching the square symmetry [e.g. (4,12), (1,4,16), etc.] but such states were never observed in our experiments. Understanding this requires detailed calculations of free energies for the states with square and circular symmetries which are not available at the moment.

In terms of the relationship between vorticity and the total flux through the disk, it was not possible to distinguish between different states with the same vorticity, as one can see in Fig. 3*b*. For example, the rarely observed state (1,8) and the stable state (2,7) are found over the same interval of $\Phi/\Phi_0$, providing another indication that it is factors such as pinning and local conditions, rather than the external field, that determine whether vortices in a particular disk achieve the lowest-energy configuration. At the same time, as Fig. 3*b* demonstrates, we could trace the evolution of vortex states as the vorticity increased with increasing flux. In agreement with theory, the states with different *L* are realised over overlapping intervals of magnetic flux and their evolution follows a well-defined pattern: As the magnetic flux through the disk increases, the first 5 vortices are added one by one to form the first circular shell – images of states (3) and (4) are shown in Fig. 3*b*, of state (5) in Fig. 4. At *L*=6, the second shell appears in the form of one vortex in the centre and this configuration remains stable until *L*=9 is reached, i.e. the next 3 vortices are added to the first (outer) shell – see open circles in Fig. 3*b* and images of states (1,6) and (1,7) in Figs. 3*b* and 4, respectively. The second (inner) shell begins to grow at *L*=9, with the next two states having 2 vortices in the centre [states (2,7) and (2,8) – see images in Fig. 3*a*], the following two states having 3 vortices in the centre [state (3,7) in Fig. 3a and (3,8) in Fig. 3*b*] and so on.

The process of shell filling for all vorticities between *L* =0 and 40 is summarised in Fig. 4. For *L*=12 to 16, the vortices are added intermittently either to the first or the second shell. At L=17 the third shell appears, again in the form of one vortex in the centre. The next 3 vortices are added to the first (outermost) shell, after which all three shells grow intermittently until the vorticity reaches *L*=32. The fourth shell appears at *L*=33 in the form of one vortex in the centre and a process identical to the evolution of the second and third shells is repeated again.

We note that vortex patterns with one vortex in the centre are particularly stable for states with more than one shell, as they are observed for several consecutive values of *L*. Indeed, this is expected to be a property of vortex numbers close to those with 'triangular' shell numbers N=1+6(1+2+3…) [15], e.g. states (1,6), (1,7), (1,6,12), (1,6,13) and so on. The stability of states (1,6) and (1,7) over relatively large intervals of external field is also evident from Fig. 3b. Furthermore, once a new shell appears, the vortex configuration never reverses to the previous number of shells and there is a fairly clear tendency that the next few vortices are added to the next but one shell (e.g., as the 4$^{th}$ shell appears for *L*=33, the next few vortices are mostly added to the 2$^{nd}$ shell). Therefore, we can unambiguously identify the vorticity corresponding to one, two, and three closed shells. We emphasise that the vorticity values corresponding to the appearance of new shells are very robust, as they were reproduced in several experiments, with different external field and disk sizes, and therefore represent 'magic' numbers for vortices confined in a circular disk [24]. For our system these numbers are 5, 16 and 32 as indicated by arrows in Fig. 4 (the first magic number, *L*=5, corresponds to the stable states (5) and (1,5), rather than the rarely observed metastable state (6) – see the discussion above).

The above pattern of shell filling for vortices is somewhat reminiscent of the formation of shells in atoms and nuclei. Unfortunately, at the moment there is no theory that would explain the physical mechanism responsible for populating each vortex shell, e.g. similar to Hund's rules. It would be very interesting to try and identify such rules, in addition to the current understanding that the appearance of each new shell is somehow dictated by achieving the lowest energy for the whole vortex configuration.

In conclusion, well-defined shell structures have been observed directly for up to 40 vortices in small circular disks. Statistical analysis of vortex configurations in hundreds of samples allowed us to identify stable and metastable states for different vorticities. Furthermore, we identified the rules of shell filling and 'magic' numbers, corresponding to the appearance of new vortex shells.

*Acknowledgements*. This work was supported by the UK Engineering and Physical Sciences Research Council.

| $L$ | $S_{exp}$ | $S_{theory}$ | $MS_{exp}$ | $MS_{theory}$ |
|---|---|---|---|---|
| 1 | 1 | 1 | - | - |
| 2 | 2 | 2 | - | - |
| 3 | 3 | 3 | - | - |
| 4 | 4 | 4 | - | - |
| 5 | 5 | 5 | (1,4) | - |
| 6 | (1,5) | (6) | (6) | (1,5) |
| 7 | (1,6) | (1,6) | - | (7) |
| 8 | (1,7) | (1,7) | - | (8) |
| 9 | (2,7) | (1,8) | (1,8); (1,8)$_\diamond$ | - |
| 10 | (2,8) | (1,9) | (3,7); (2,8)$_\diamond$ | (2,8) |
| 11 | (3,8) | (2,9) | (4,7) | (1,10); (3,8) |

TABLE I. Comparison of the experimentally observed stable ($S_{exp}$) and metastable ($MS_{exp}$) states for $L \leq 11$ with those found numerically in ref. [6] ($S_{theory}$ and $MS_{theory}$, respectively).

FIGURE CAPTIONS

FIG. 1. SEM images of vortex patterns observed in the starting macroscopic film (*a*) and in mesoscopic dots (*b,c*). A few vortices in each dot can just be discerned in (*b*), while (*c*) shows a higher-magnification view of vortex configurations in disks with $d \approx 2$, 3 and 5μm obtained in the same experiment at $H =40$ Oe. The scale bar on all images is 5μm.

FIG. 2. Vorticity *L* as a function of the normalised magnetic flux through the disk area, $\Phi/\Phi_0$. Different symbols correspond to different experiments. Straight solid line shows the corresponding dependence for the same area of a macroscopic film. The inset shows normalised excess flux $\delta\Phi/\Phi_0=(\Phi-\Phi_0 L)/\Phi_0 L$ corresponding to adding one vortex to a state (*L*-1) in order to form a state *L*. The dashed line shows a fit to the function $\delta\Phi/\Phi_0=(a+bL)/(1+cL)$ found in ref. [5], with fitting coefficients *a*=4.75, *b*=0.26 and *c*=1.2. A corresponding fit for $\Phi/\Phi_0$ vs *L* is shown by the dashed line in the main panel.

FIG. 3. (a) Histogram showing the distribution of different vortex states observed in the same experiment ($H_{ext} =60$ Oe) on disks with diameter $d \approx 3$μm. SEM images of the corresponding states are shown as insets. (b) Evolution of vortex states with increasing flux through the sample (arrows are guide to the eye). Different symbols correspond to different shell configurations: ◆ - represent states (*L*), ○ - states (1, *L*-1), ● - states (2, *L*-2) and ◇ - states (3, *L*-3). SEM images of some of the states are shown as insets: [left to right] states (3); (4); (6); (1,6); (3,8).

FIG. 4. Number of vortices populating different shells as a function of vorticity *L*. All configurations observed in different external fields and on disks of different sizes are included. Two or more different states were found for some vorticities (see text), e.g. (1,7,14) and (2,7,13) for *L*=22; (1,5,12,17), (1,5,11,18), (1,6,11,17) and (1,6,12,16) for *L*=35, etc. Open circles show vortex numbers corresponding to the states that we could identify as metastable. Arrows indicate 'magic' numbers, i.e., vorticity values corresponding to closed shell configurations, just before the appearance of the $2^{nd}$, $3^{rd}$ and $4^{th}$ shell. SEM images of vortex states with a corresponding number of shells are shown as insets.

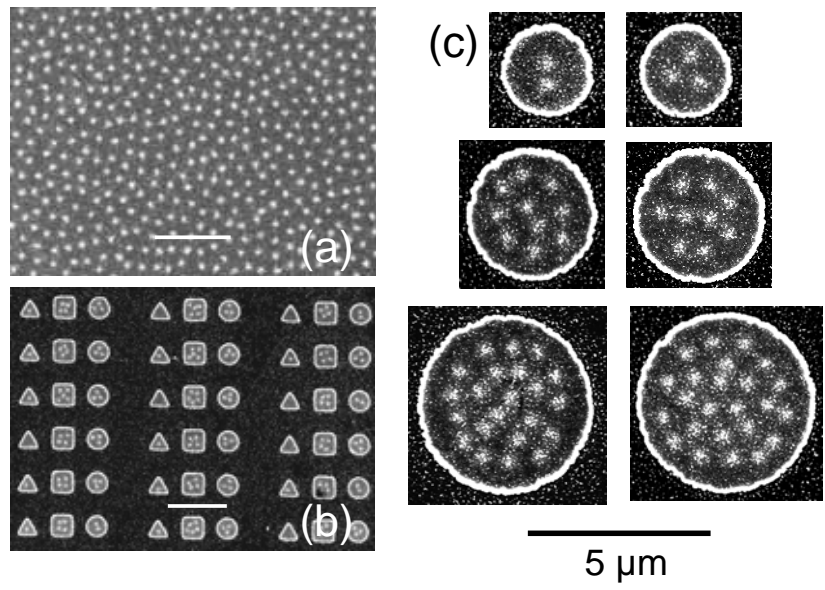

Figure 1

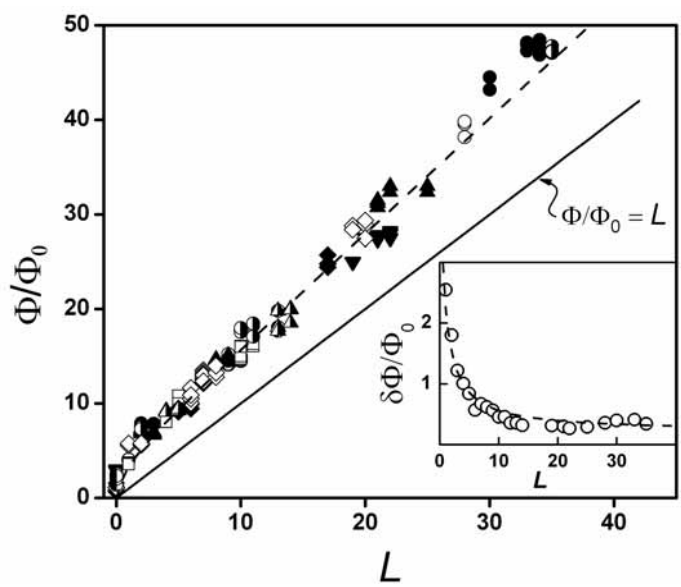

Figure 2

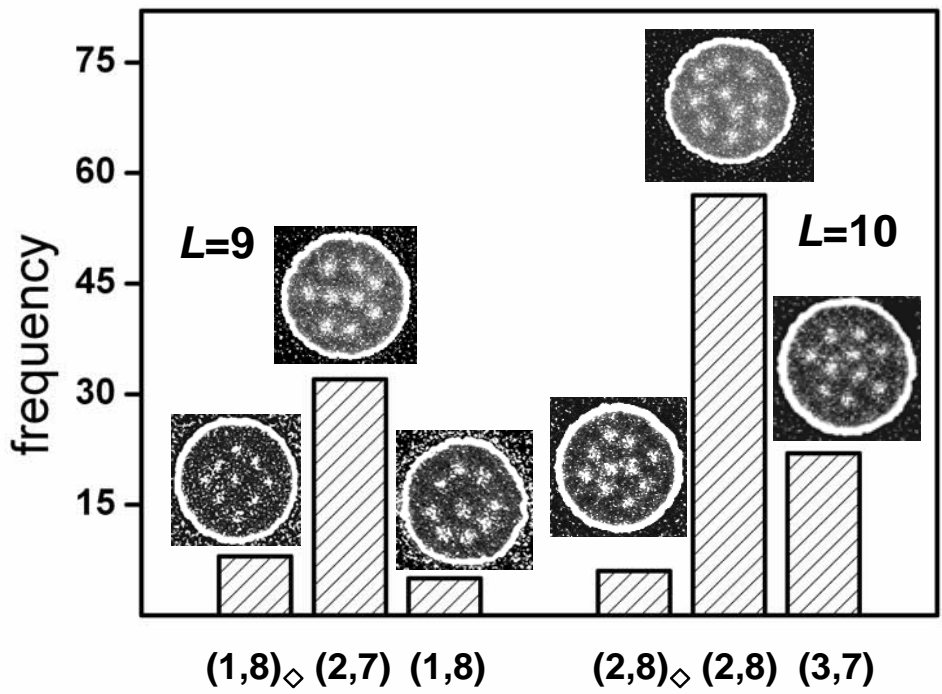

Figure 3a

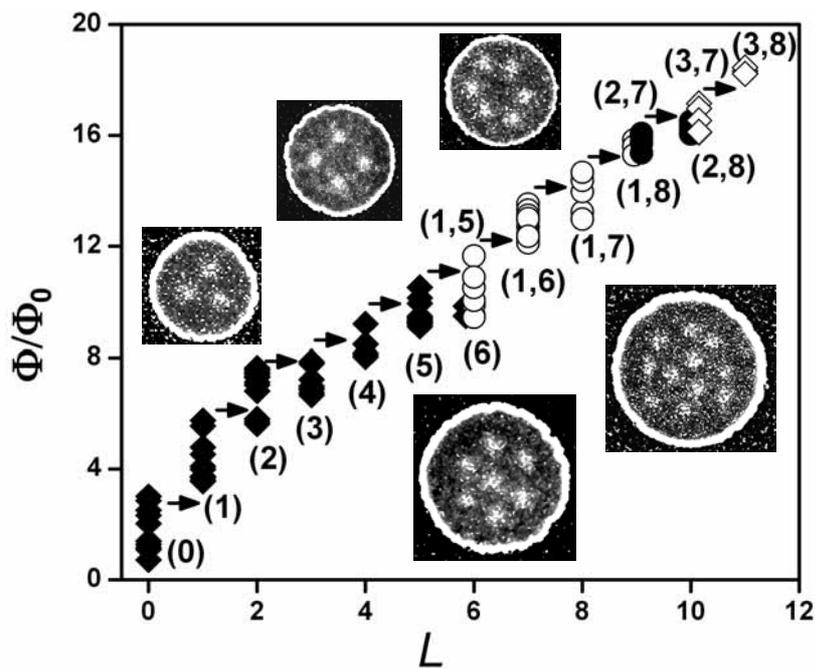

Figure 3b

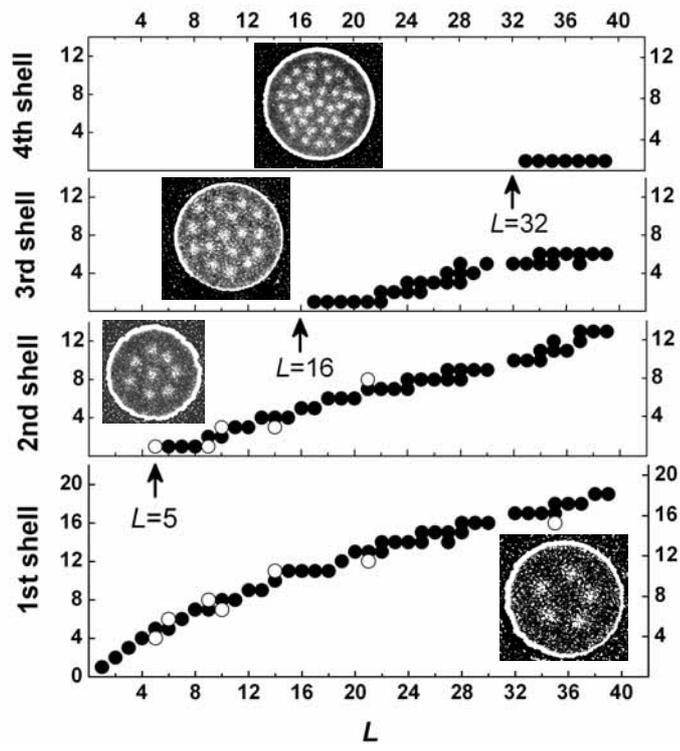

Figure 4